\documentclass[aps,prl,twocolumn,reprint,amsmath,amssymb,superscriptaddress,floatfix,footinbib,longbibliography]{revtex4-1}

\usepackage{blindtext}
\usepackage{epsfig}
\usepackage{graphicx}
\usepackage{epstopdf}
\usepackage[T1]{fontenc}
\usepackage[applemac]{inputenc}
\usepackage{lmodern}
\usepackage[english]{babel}
\usepackage{ae}
\usepackage{units}
\usepackage{amsmath,amssymb,natbib,bm}
\usepackage{psfrag}
\usepackage{subfigure}
\usepackage{amsthm}
\usepackage{bbm}
\usepackage{booktabs}
\usepackage[americaninductors]{circuitikz}
\usepackage{tikz}
\usetikzlibrary{arrows}
\usepackage{color}
\usepackage{url}
\usepackage[colorlinks]{hyperref}
\hypersetup{%
        plainpages=true,
        breaklinks=true,
        hypertexnames=false,
        pageanchor=true,
        colorlinks=true,
        linkcolor={blue},
        citecolor={magenta},
        urlcolor={blue},
        anchorcolor={black}
      }
\usepackage[all]{hypcap} 
\usepackage{mleftright} 
\usepackage{longtable}
\usepackage[normalem]{ulem}

\newcommand{\figref}[1]{\mbox{Fig.~\ref{#1}}}
\newcommand{\tabref}[1]{\mbox{Table~\ref{#1}}}

\renewcommand{\eqref}[1]{\mbox{Eq.~(\ref{#1})}}
\newcommand{\figpanel}[2]{Fig.~\hyperref[#1]{\ref*{#1}(#2)}}
\newcommand{\figpanels}[3]{Fig.~\hyperref[#1]{\ref*{#1}(#2)-(#3)}}
\newcommand{\figpanelNoPrefix}[2]{\hyperref[#1]{\ref*{#1}(#2)}}

\newcommand{\ket}[1]{|#1\rangle}

\newcommand{\ketbra}[2]{\mleft| #1 \rangle \langle #2 \mright|}

\newcommand{\tr}[1]{\text{tr}\mleft( #1 \mright)}

\newcommand{\im}{\text{Im}}
\newcommand{\re}{\text{Re}}

\newcommand{\be}{\begin{equation}}
\newcommand{\ee}{\end{equation}}
\newcommand{\bea}{\begin{eqnarray}}
\newcommand{\eea}{\end{eqnarray}}

\AtBeginDocument{%
    \newwrite\bibnotes
    \def\bibnotesext{Notes.bib}
    \immediate\openout\bibnotes=\jobname\bibnotesext
    \immediate\write\bibnotes{@CONTROL{REVTEX41Control}}
    \immediate\write\bibnotes{@CONTROL{%
    apsrev41Control,author="08",editor="1",pages="0",title="0",year="1"}}
     \if@filesw
     \immediate\write\@auxout{\string\citation{apsrev41Control}}%
    \fi
}%

\begin{document}


\title{Gradient-descent quantum process tomography by learning Kraus operators}

\author{Shahnawaz Ahmed}
\email{shahnawaz.ahmed95@gmail.com}
\affiliation{Department of Microtechnology and Nanoscience, Chalmers University of Technology, 412 96 Gothenburg, Sweden}

\author{Fernando Quijandr\'{i}a}
\affiliation{Department of Microtechnology and Nanoscience, Chalmers University of Technology, 412 96 Gothenburg, Sweden}
\affiliation{Quantum Machines Unit, Okinawa Institute of Science and Technology Graduate University, Onna-son, Okinawa 904-0495, Japan}

\author{Anton Frisk Kockum}
\email{anton.frisk.kockum@chalmers.se}
\affiliation{Department of Microtechnology and Nanoscience, Chalmers University of Technology, 412 96 Gothenburg, Sweden}

\date{\today}


\begin{abstract}

We perform quantum process tomography (QPT) for both discrete- and continuous-variable quantum systems by learning a process representation using Kraus operators. The Kraus form ensures that the reconstructed process is completely positive. To make the process trace-preserving, we use a constrained gradient-descent (GD) approach on the so-called Stiefel manifold during optimization to obtain the Kraus operators. Our ansatz uses a few Kraus operators to avoid direct estimation of large process matrices, e.g., the Choi matrix, for low-rank quantum processes. The GD-QPT matches the performance of both compressed-sensing (CS) and projected least-squares (PLS) QPT in benchmarks with two-qubit random processes, but shines by combining the best features of these two methods. Similar to CS (but unlike PLS), GD-QPT can reconstruct a process from just a small number of random measurements, and similar to PLS (but unlike CS) it also works for larger system sizes, up to at least five qubits. We envisage that the data-driven approach of GD-QPT can become a practical tool that greatly reduces the cost and computational effort for QPT in intermediate-scale quantum systems.

\end{abstract}

\maketitle


\paragraph*{Introduction.}

The characterization of quantum operations in the presence of various coherent and incoherent noise sources is of broad interest in emerging quantum technologies~\cite{Feynman1982, Lloyd2008, Kimble2008, You2011, Bal2012, Georgescu2014, Montanaro2016, Gu2017, Kandala2017, Wendin2017, Degen2017, Pirandola2018, Preskill2018, Childs2018, Arguello-Luengo2019, Gefen2019, Korobko2019, Arute2019, Ma2020, McArdle2020, Bauer2020, Barzanjeh2020, Wang2020, Yin2020, Zhong2020, Cerezo2021, Madsen2022}. Some of the quantum operations involved are, e.g., implementing and sampling from the output of a complicated quantum circuit~\cite{Arute2019, Zhong2020} or creating interesting quantum states~\cite{Grimm2020, Campagne-Ibarcq2020, Kim2021, Kudra2022}. The ability to characterize such operations efficiently would enable tackling the noise and understanding the limitations of current quantum devices.

A quantum operation can be generally represented as a completely positive (CP) and trace-preserving (TP) linear map $\mathcal E$ mapping a quantum state $\rho$ in a Hilbert space to another state $\rho'$ in a (possibly) different Hilbert space, i.e., $\rho' = \mathcal E(\rho)$. The task of estimating $\mathcal E$ from experimental data is called quantum process tomography (QPT)~\cite{Poyatos1997, Chuang1997, Ariano2001, Fedorov2015}. It can be performed using techniques such as maximum-likelihood estimation~\cite{Fiurasek2001, Sacchi2001, Rahimi-Keshari2011, Anis2012}, Bayesian estimation~\cite{Schultz2019}, projected gradient descent (GD)~\cite{Knee2018}, projected least-squares (PLS)~\cite{Surawystepney2021}, compressed-sensing (CS) methods using convex optimization~\cite{Baldwin2014, Rodionov2014, Banchi2020, Teo2020}, or emerging ideas like variational QPT~\cite{Xue2022}.

One of the main challenges in QPT is that the size of the process representation grows exponentially with the size of the quantum system. For example, the Choi representation~\cite{Choi1975} is a complex-valued $4^n \times 4^n$ matrix for $n$ qubits. This makes it difficult both to estimate a quantum process from noisy data, and to interpret the results~\cite{Kofman2009}. However, for realistic cases, low-rank approximations of a quantum process require much less data for QPT~\cite{Rodionov2014}. In the case of quantum state tomography (QST), such approximations are sufficient to guide toward actionable and interpretable information~\cite{Riofrio2017}. Therefore, inspired by machine-learning techniques such as neural-network QST~\cite{Carleo2017, Carleo2018, Torlai2018, Carrasquilla2019, Melkani2019, Torlai2020, Smith2020, Lohani2020, Neugebauer2020, Palmieri2020, Banchi2018, Che2021}, efficiently representing a process and learning it from data has recently shown promise for QPT~\cite{Torlai2020}. Other approaches, e.g., shadow tomography~\cite{Levy2021, Kunjummen2021}, circumvent the problem of large Hilbert-space dimension by completely avoiding the construction of a process representation.

In this Letter, we tackle the QPT problem using a simple GD-based optimization that learns the Kraus representation of a process. Kraus operators can represent any completely positive linear map~\cite{Choi1975, Leung2003}. The number of Kraus terms can be flexibly adapted for a low-rank reconstruction. We treat the process estimation as a learning task performed by using batches of measurement data, similar to the training of neural networks. The CP condition is ensured by construction~\cite{Leung2003} and an efficient gradient-retraction method~\cite{Wen2013, Jiang2015, Li2020, Adhikary2020} implements orthonormality constraints such that the TP condition is never violated during the optimization.

Our GD-QPT approach is an example of Riemannian optimization on the Stiefel manifold~\cite{Tagare2011, Boumal2022}, which is also of interest in machine learning~\cite{Li2020}, and has been applied to problems in quantum physics~\cite{Adhikary2020, Wiersema2022}. We show how this simple approach can be used to reconstruct both continuous- and discrete-variable (CV/DV) quantum processes with different measurement schemes, large Hilbert-space dimensions, and a small amount of data.

We benchmark our approach against CS and PLS. In CS, the Choi representation is used to turn the QPT task into a convex optimization problem~\cite{Baldwin2014, Rodionov2014, Teo2020}. Using convex programming, CS can handle CPTP constraints easily, guarantees a global optimum, and shines in the regime where the amount of data is so small that standard QPT would have an underdetermined system of equations for a general full-rank process~\cite{Rodionov2014}. However, CS can be computationally expensive in practice, limiting its applicability for even three-qubit processes, which may require several hours of computation~\cite{Rodionov2014}. A five-qubit process, or a CV process with a larger cutoff on the Hilbert space than used in existing methods~\cite{Rahimi-Keshari2011, Anis2012}, may thus become impractical with larger Choi matrices and more data.

Projected-gradient methods~\cite{Knee2018} like PLS~\cite{Surawystepney2021} first obtain an estimate for the process and then project it to the nearest CPTP estimate. The PLS method is fast: it can reconstruct processes for 5--7 qubits in a reasonable amount of time~\cite{Surawystepney2021}. However, projection-based techniques may require finding an initial analytical least-squares estimate relying on an informationally complete set of measurements, along with costly projection steps involving eigendecompositions of large Choi matrices or iterative subroutines. Our approach avoids these problems while still being able to handle relatively large Hilbert space dimensions. 

We show that the simple GD-QPT technique yields similar performances as CS and PLS on benchmarks using random processes with Gaussian noise in the data. We also assess the performance of GD-QPT and CS against the amount of data, to show that GD-QPT combines the best of two worlds. Like CS, the GD-QPT approach works with a less than informationally complete set of measurements, but it can still, like PLS, be run for larger problems. 

As an extension to GD-QPT, we also try out neural-network QPT (NN-QPT)~\cite{SuppMat}. In NN-QPT, the Kraus operators are given by the output of a neural network, similar to ideas explored in previous works~\cite{Ahmed2020a, Ahmed2020b} for QST. We find no significant advantage of this approach compared to GD-QPT, which indicates that a good process representation, along with constrained GD optimization, might be sufficient to learn quantum states and processes. Our approach thus introduces a flexible QPT technique, demonstrating that simple gradient-based optimization combined with appropriate regularization and efficient process representation can become an effective tool for quantum process characterization.


\paragraph*{Kraus and Choi representations.}

The Kraus-operator representation of $\mathcal{E}$ corresponds to $k$ complex-valued matrices \{$K_l$\} of dimension $N \times N$, that act on a density matrix $\rho$ as $\mathcal E(\rho) = \sum_{l = 1}^{k} {K_l} \rho {K_l}^{\dagger} = \rho'$. Here $N$ is the Hilbert-space dimension; for $n$ qubits $N = 2^n$. The Kraus representation guarantees that the process is CP~\cite{Leung2003, Bhatia2009, Kadri2020}. The TP condition translates to the Kraus operators satisfying $\sum_{l = 1}^{k} K_l^{\dagger} K_l = \mathbb{I}$.

The Choi representation~\cite{Choi1975, Jamiolkowski1972} of $\mathcal{E}$ is a single $N^2 \times N^2$ complex-valued matrix $\Phi$ that can be written using Kraus operators as $\Phi = \sum_{l = 1}^{k} \vert K_l \rangle \langle {K_l}\vert$ with $\vert K_l \rangle = (\mathbb{I} \otimes K_l) \sum_i \vert i \rangle \otimes \vert i \rangle $. The Choi matrix is thus a linear operator acting on the tensor product of the input and output Hilbert spaces $H_{\rm in} \otimes H_{\rm out}$. The action of $\Phi$ on a state $\rho$ is given by the partial trace operation $\rho' = \text{Tr}_{H_{\rm in}}[(\rho^{T} \otimes \mathbb I)\Phi]$. In order for $\Phi$ to be CPTP, it should be positive-semidefinite~\cite{Choi1975, Knee2018} and satisfy $\text{Tr}_{H_{\rm out}}(\Phi) = \mathbb I$.

In our approach to QPT, we consider the Kraus-operator form because it allows us to control the size of the process representation. The Choi rank $r$ of a process is given by the minimum number of Kraus operators necessary to represent the process. The maximum Choi rank is $r = N^2$, but in realistic cases, process matrices can have low ranks $r \ll N^2$ (for a unitary process, $r = 1$). Our approach thus gives the flexibility to choose the rank $r = k$ (the number of Kraus operators) of the process ansatz, allowing us to obtain low-rank approximations without constructing the full Choi matrix. In most previous QPT methods, the Choi-matrix representation was preferred because it made CPTP constraints easier to handle~\cite{Knee2018} and the problem could be cast in a linear form~\cite{Rodionov2014}.


\paragraph*{Learning quantum process representations.}

\begin{figure}
    \centering
    \includegraphics[width=\columnwidth]{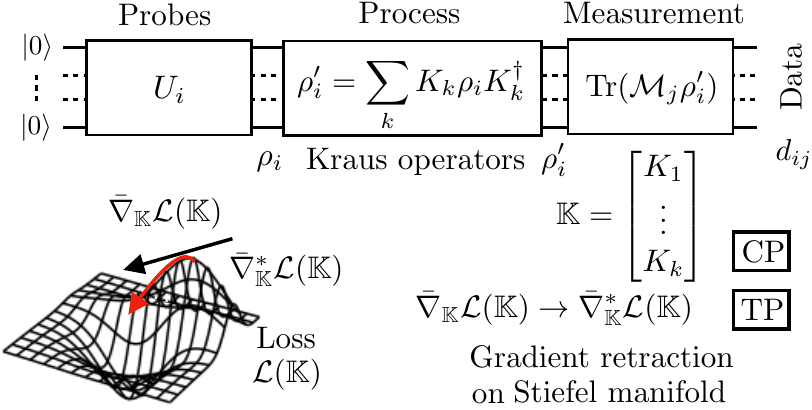}
    \caption{An illustration of GD-QPT. In QPT, we estimate a quantum process $\mathcal E$ from data $d_{ij}$ that represent expectation values of measurements $\mathcal M_j$ on states given by $\mathcal E$ acting on probes $\rho_i$. We construct a loss function $\mathcal L$ using the data and our estimate for the process as a set of Kraus operators stacked together, $\mathbb K$. We estimate $\mathbb K$ by minimizing the loss with GD-based optimization using batches of data. The Kraus form guarantees that the estimate is CP and a gradient-retraction technique restricts each GD update such that $\mathbb K$ remains in the set of orthonormal matrices on the Stiefel manifold, ensuring that it is also TP.
    \label{fig:qpt}}
\end{figure}

We illustrate the idea of GD-QPT in \figref{fig:qpt}. In QPT experiments, data $d_{ij}$ are obtained as estimates (computed by averaging single-shot outcomes) of expectation values of measurements $\mathcal M_j$ on the output states $\rho_i' = \mathcal E(\rho_i)$ for given probe (input) states $\rho_i$. We model statistical errors in the measurement as noise sampled from a zero-mean Gaussian distribution $\mathcal N(0, \epsilon)$ with standard deviation $\epsilon$. The process-reconstruction problem can then be cast as a learning task: minimizing a loss function $\mathcal L$ representing the discrepancy between the data $d_{ij}$ and our estimate of the process.
We use
\be
    \label{eq:loss}
    \mathcal L(\mathbb K) = \sum_{ij}\mleft[d_{ij} - \text{Tr}\mleft[\mathcal M_j  \mleft(\sum_k {K_k} \rho_i {K_k}^{\dagger}\mright)\mright]\mright]^2 + \lambda \vert \vert \mathbb K\vert \vert_{1},
\ee
which combines least-squares-error loss with L1 regularization. Here, $\mathbb K = [K_1, \dots, K_k]$ is a $kN \times N$ matrix, which represents the process by stacking the $k$ Kraus operators, and $\vert \vert \mathbb K\vert \vert_{1}$ is given by the L1 norm~\cite{Santosa1986, Tibshirani1996}, with $\lambda \ge 0$ the strength of the regularization. We fix $\lambda = 10^{-3}$ in this Letter, but it can be optimized further as a hyperparameter.

The loss function can be minimized with GD by updating $\mathbb K$ along the negative (conjugate, since the Kraus operators are complex~\cite{Hjorungnes2007}) gradient $\bar \nabla_{\mathbb K} \mathcal L(\mathbb K)$. However, simple GD might lead to an estimate that violates the TP constraint, which can be equivalently described as the orthonormality condition $\mathbb K^{\dagger} \mathbb K = \mathbb I$. To counter such violations, one could add a loss term that penalizes them, e.g., $\vert \vert \sum_l K_l^{\dagger} K_l - \mathbb{I} \vert \vert_1$. However, this penalty does not strictly enforce the orthonormal condition.

In the Choi representation, we can linearize the problem to implement CS-QPT as
\be
    \Phi_{\text{CS}} = \text{arg\,min} ||\Phi'||_1 \,\, \text{s.t.} \, \Phi' \ge 0 , \vert \vert S\vec \Phi' - \vec d\vert \vert_2 \le \delta ,
\ee
where $\delta$ is the noise level we set as a threshold. The matrix $S$ is similar to the sensing matrix in QST~\cite{Shen2016}, which is constructed using the probes and measurement operators $\{\rho_i, \mathcal M_j\}$~\cite{Knee2018}. The data is collected into a vector $\vec d$ with an appropriate flattening $\vec \Phi'$ of the Choi matrix. The TP condition is implemented by setting the constraint $\text{Tr}_{H_{\rm out}}(\Phi') - \mathbb I = 0$. We use the splitting conic solver~\cite{ODonoghue2016} to solve the convex optimization task for CS in Python with CVXPY~\cite{diamond2016cvxpy, agrawal2018rewriting} following Qiskit~\cite{Qiskit2021} to implement the CPTP constraints.


\paragraph*{Gradient descent on the Stiefel manifold.}

The orthonormal condition on $\mathbb K$ defines the so-called Stiefel manifold~\cite{Tagare2011}. It is possible to restrict the gradients such that we never leave this manifold during the optimization~\cite{Tagare2011, Wen2013, Jiang2015, Li2020, Boumal2022}; this is an example of Riemannian optimization on a manifold~\cite{Boumal2022}. Several works have addressed this problem using a so-called retraction technique that is an approximation to the exponential map~\cite{Absil2009}. The retraction restricts the updated $\mathbb K$ to the Stiefel manifold while minimizing the loss (see \figref{fig:qpt}).

Let $G'=\bar \nabla_{\mathbb K} \mathcal{L}(\mathbb{K})$. At each update step, we normalize the gradients with the L2 norm as $G = G'/||G'||_2$. If $A = [G \,\,\,\,\,\,\mathbb{K}]$ and $B = [\mathbb K \,\,-G]$, representing stacked matrices, the trace-preserving retraction is given by
\be
\bar \nabla_{\mathbb K}^{*}\mathcal L(\mathbb K) = A(\mathbb I + \frac{\eta}{2}B^{\dagger}A)^{-1} B^{\dagger}\mathbb K,
\ee
where $\eta$ is a learning rate, such that we can iteratively apply the gradient updates $\mathbb K' = \mathbb K - \eta \bar \nabla_{\mathbb K}^{*}\mathcal L(\mathbb K)$ to minimize the loss $\mathcal L(\mathbb K)$ while keeping $\mathbb K$ in the Stiefel manifold. The retraction formula is based on the Cayley transform and the use of the Sherman-Morrison-Woodbury formula~\cite{Tagare2011}.

The starting estimate for the Kraus operators are taken to be random unitary matrices with appropriate normalization guaranteeing that they are CPTP. We consider a learning rate that decays by a factor 0.999 in each step with $\eta (0) = 0.1$.


\paragraph{Results and benchmarking.}


\begin{figure}
    \centering
    \includegraphics[width=\columnwidth]{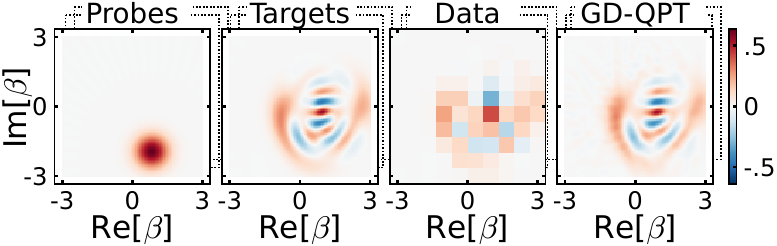}
    \caption{Applying gradient-based learning of Kraus operators to reconstruct a CV quantum process with a Hilbert-space cutoff of $32$. The probe states are coherent states $\ket{\alpha_i}$ in a $10 \times 10$ grid with $\re[\alpha_i], \im[\alpha_i] \in [-2.5, 2.5]$. The measurements are the value of the displaced parity $\Pi(\beta_j)$ in a $10 \times 10$ grid with $\re[\beta_j], \im[\beta_j] \in [-3, 3]$. The data collected is chosen to be very coarse, to demonstrate that we only need a few measurements for each probe in our reconstruction. We sample more data in a finer grid ($d_{ij}'$) from our process estimate to show the match with the true process.
    \label{fig:snaps}}
\end{figure}

We start with an example of QPT for a CV quantum operation --- a selective number-dependent arbitrary phase (SNAP) gate~\cite{Heeres2015, Fosel2020, SuppMat} along with a displacement operation --- using coherent states as probes~\cite{Rahimi-Keshari2011}. The SNAP gate has been recently used experimentally to create a variety of interesting CV quantum states such as Gottesman--Kitaev--Preskill states and the cubic phase state~\cite{Kudra2022}. The parameters of the chosen SNAP + displacement operation are given in the Supplementary Material~\cite{SuppMat}. In such CV problems, choosing an appropriate Hilbert-space cutoff, which allows to correctly describe the state at hand, is fundamental~\cite{Rahimi-Keshari2011, Anis2012}. Here, we consider a cutoff of $32$, which, to the best of our knowledge, is the largest dimension explored for single-mode CV QPT~\cite{Rahimi-Keshari2011, Kumar2013, Fedorov2015, Kupchak2015}.

In \figref{fig:snaps}, we show the Wigner functions for a single instance of a coherent probe state $\rho_i = \ketbra{\alpha_i}{\alpha_i}$, target state $\rho_i' = \mathcal E(\rho_i)$ after the process, and sampled data $d_{ij}$. The data corresponds to measurements of the displaced parity operator $\Pi(\beta)$ on $\rho_i'$. Since both $\alpha_i$ and  $\beta$ are continuous, we need to select a grid to run an experiment. We deliberately choose a very coarse grid to highlight that we do not require full Wigner tomography for each probe state during QPT. In an experiment, appropriate choices have to be considered for the probes and measurements depending on the process as well as the limitations of the experimental setup.

There are several distance measures available to quantify the difference between quantum processes~\cite{Gilchrist2005, Nechita2018}. We consider the fidelity $F (\Phi, \Phi') = \tr{\sqrt{\sqrt{\Phi} \Phi' \sqrt{\Phi}}}$ of reconstruction according to the definition of process fidelity~\cite{Wood2015} by converting our Kraus reconstruction to the Choi form with the appropriate normalization, $\Phi^{\text{GD-QPT}}/N$. The average value of $F$ for $30$ random choices of the Kraus operators with a noise $\epsilon = 10^{-2}$ is $>0.97$. Each reconstruction converges within $50$ iterations taking tens of seconds on a standard laptop with a \unit[32]{GB} memory and \unit[2.9]{GHz} 6-Core Intel Core i9 processor. We used $k=3$ Kraus terms in the reconstruction even though we actually only needed to have $k=1$ since the SNAP and displacement operations are unitary.


\begin{figure}
    \centering
    \includegraphics[width=\columnwidth]{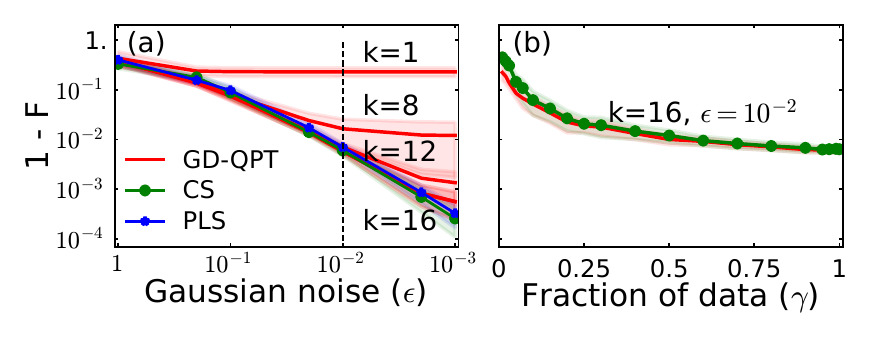}
    \caption{Benchmarking GD-QPT (red) against PLS (blue) and CS (green) for random two-qubit ($n=2$) full-rank ($r=16$) processes. (a) Mean infidelities for various number of Kraus operators as the Gaussian noise level $\epsilon$ is decreased for $30$ random processes. Shading shows one standard deviation. (b) Average infidelities for reconstructions of the same processes using a fraction $\gamma$ of the total data used in (a). We randomly select a $\sqrt \gamma/6^n \times \sqrt \gamma/6^n$ subset of the total $6^n \times 6^n$ Pauli probes and measurements.
    \label{fig:benchmark}}
\end{figure}

To study the effects of the number of Kraus operators in our ansatz, noise, and amount of data, we turn to the reconstruction of random DV quantum processes, i.e., processes acting on $n$ qubits. First, in \figpanel{fig:benchmark}{a}, we quantify the impact of Gaussian measurement noise (related to the number of measurement samples~\cite{Surawystepney2021}) and the number of Kraus operators. We follow the direct QPT approach of Ref.~\cite{Surawystepney2021}, where the $6^n$ probes and $6^n$ measurements are given by the tensor products of the eigenstates to the Pauli matrices $\{\sigma_{x}, \sigma_{y}, \sigma_{z}\}$. We compare our results on DV QPT against the  hyperplane-intersection projection method of~\cite{Surawystepney2021} and a CS implementation using convex programming~\cite{Rodionov2014}.

We find that all three methods perform similarly as a function of the measurement noise $\epsilon$ with the assumption of a full set of $k=4^n$ Kraus operators for a full-rank process. As we reduce the number of Kraus terms in our ansatz, the fidelity saturates at lower values for lower $k$s, which is expected as our approximation is not expressive enough to represent the full-rank process with fewer Kraus terms. In a practical setting, we may only be able to reach a certain fidelity with a finite number of measurement shots (non-zero $\epsilon$). Interestingly, using more Kraus terms in such situations may not be helpful.

In most realistic cases, where we might be interested in implementing quantum gates which are unitary or near-unitary processes, CS-QPT methods can work with very little data~\cite{Rodionov2014}. Nevertheless, we benchmark for full-rank processes in a two-qubit system to demonstrate the general applicability of our approach. In \figpanel{fig:benchmark}{b}, we compare the performance of GD-QPT against CS with a fixed noise level $\epsilon = 10^{-2}$ for full-rank ($k=16$) Kraus operators using a random subset of probes and measurements (a fraction $\gamma$ of the total). We observe that for \textit{two-qubit} processes in this informationally incomplete regime, GD-QPT achieves similar performance as CS --- both needing less than half of the total data. In the Supplementary Material~\cite{SuppMat}, we also present results for low-rank processes. By using a smaller number of Kraus operators, we can reconstruct processes in a larger Hilbert space (5-qubit DV systems and CV processes with a Hilbert-space cutoff of $32$).

There could be further possible improvements with lower noise (higher number of shots) and tuning of the hyperparameters such as regularization, learning rate, and number of Kraus operators. It is also important to note that for such reconstructions, assessing the uncertainty of the reconstruction becomes important since we cannot guarantee if we have enough data for a unique estimation of the process~\cite{Kiktenko2021}.

The PLS method was omitted from \figpanel{fig:benchmark}{b} since it was not clear how to adapt it to non-informationally complete data. However, PLS can be used to reconstruct processes with more qubits, where CS has difficulties. The dimension of the matrix $S$ in CS is $6^{2n} \times 4^{2n}$ for the probabilities and the flattened Choi matrix. Therefore, running convex optimization programs even for a three-qubit process reconstruction with CS is challenging, requiring several hours of computational time~\cite{Rodionov2014}. In contrast, GD-QPT can easily tackle five-qubit processes, similar to PLS. Further, due to the restricted number of Kraus operators, GD-QPT iterations are faster than PLS for larger Hilbert spaces. 

\begin{figure}
    \centering
    \includegraphics[width=\columnwidth]{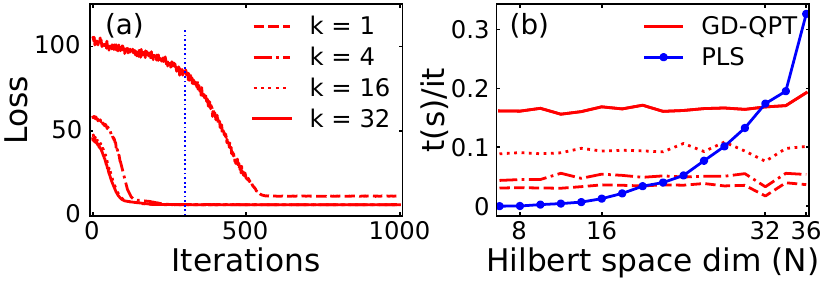}
    \caption{Comparing computational time for GD-QPT (red) and PLS (blue). (a) Loss in GD-QPT after each update step for random 5-qubit DV processes of rank $r=3$, using various numbers of Kraus terms. Each such step for GD-QPT takes a batch of $256$ data points from a total of $6^{5} \times 6^{5}$ expectation values, computes the loss and its gradient, and performs an update on the Stiefel manifold. The processes have rank $r=3$. (b) Time taken for each GD-QPT iteration and the most expensive step in PLS, the CP projection, as a function of Hilbert-space dimension with random processes~\cite{SuppMat}. In Ref.~\cite{Surawystepney2021}, around $300$ iterations were needed for convergence of the CP projection when $n=5$.
    \label{fig:qpt-convergence}}
\end{figure}

In \figref{fig:qpt-convergence}, we compare the number of iterations for the convergence of GD-QPT against PLS for random 5-qubit DV processes. The results show that GD-QPT converges in a similar number of iterations as PLS, but is faster per iteration due to the smaller number of Kraus terms considered. The most expensive step in the PLS technique, the CP projection, requires a diagonalization involving the eigendecomposition of the $4^n \times 4^n$-dimensional Choi-matrix estimate. The time taken for each step is limited by the complexity (cubic) of this eigendecomposition. In comparison, the most expensive step in GD-QPT is the retraction involving the inversion of smaller matrices of dimensions $k 2^n \times 2^n$, where the number of Kraus terms $k \ll 4^n$. We provide the data and code for all results along with our implementation of GD-QPT, PLS, and CS in Ref.~\cite{gd-qpt2022}.

\paragraph{Conclusion and outlook.}

In this Letter, we introduced a simple yet powerful technique for QPT using gradient-based learning of Kraus operators --- GD-QPT. Our approach can reconstruct both CV and DV processes for Hilbert spaces of dimension at least $32$. We benchmarked GD-QPT against the recently proposed PLS algorithm as well as CS. Using randomly generated processes, we showed that for low-rank processes, estimating Kraus operators directly gives fidelities similar to PLS and CS for the same amount of Gaussian noise in the data.

Further, we showed that our approach performs similar to CS in the regime of informationally incomplete data, yet works for a larger number of qubits than CS. In this regard, we achieve the goal outlined in Ref.~\cite{Rodionov2014}, where the convex-programming-based CS-QPT technique suffered from numerical time and memory complexity issues. Our simple approach alleviates some of the numerical issues by considering QPT as a learning problem with a limited number of Kraus operators as a model. Using data, we learn the Kraus operators through gradient-based optimization on the Stiefel manifold.

Since the size of the Kraus operators scales exponentially with the system size, a future direction of work could be to replace the Kraus representation with other efficient (approximate) models for a quantum process, e.g., tensor networks~\cite{Torlai2020}. Benchmarking against techniques such as shadow tomography~\cite{Kunjummen2021, Levy2021} could reveal if this approach can strike a balance between tackling QPT for larger quantum systems and having an explicit, interpretable representation for the process.

It would be interesting to understand why the gradient-based approach works with so few data points and noise in the data. In this regard, the quantum-inspired low-rank Kraus decomposition technique partial trace regression~\cite{Kadri2020} can be relevant, as it shows that low-rank reconstructions are possible with a small data set. A further, crucial direction for future work is to formulate fast uncertainty estimation techniques for the reconstruction similar to existing ideas~\cite{DiMatteo2020, Kiktenko2021}. The role of both aleatoric (statistical) and epistemic (lack of data) uncertainties can be further studied to improve on the reconstruction and guide real experiments~\cite{Hullermeier2021}.


\begin{acknowledgments}

\paragraph*{Acknowledgements.}
We acknowledge useful discussions with Ingrid Strandberg, Christopher Warren, Axel Eriksson, Mikael Kervinen, Roeland Wiersema, Juan Carrasquilla, and Nathan Killoran. The Python-based packages JAX~\cite{Jax2018} and Optax~\cite{Optax2020} were used for automatic gradient calculation and optimization. QuTiP~\cite{Johansson2012, Johansson2013} was used for visualization and generating data. The CS algorithm was implemented using CVXPY~\cite{diamond2016cvxpy, agrawal2018rewriting}. We acknowledge support from the Knut and Alice Wallenberg Foundation through the Wallenberg Centre for Quantum Technology (WACQT).

\end{acknowledgments}

\bibliography{ref}

\clearpage
\appendix
\section{SUPPLEMENTARY MATERIAL}

\subsection{Definitions for CV probes, gates, and measurements}
\label{app:cv}

The CV probes used in this work are coherent states $\ket{\alpha}$ that can be defined in the Fock basis as
\be
\ket{\alpha} = D(\alpha) \ket 0,
\ee
where $D(\alpha) = \exp({\alpha a^\dagger - \alpha^* a})$ is the displacement operator with $a$ ($a^\dag$) the bosonic annihilation (creation) operator and $\ket{0}$ is the vacuum state.

The measured observable corresponds to the displaced parity operator
\be
\Pi(\beta) = \sum_n (-1)^n D(\beta) \ketbra{n}{n} D(-\beta).
\ee

The SNAP gate is defined as an operator that adds the real phase $\theta_n$ to the Fock state $\ket{n}$:
\be
S(\vec \theta) = \sum_n e^{i \theta_n} \ketbra{n}{n}
\ee
with $\vec \theta = \{\theta_n\}$ the real-valued phases that are added to each Fock state. The CV process that we consider as an example corresponds to the unitary $D(\alpha) S(\vec \theta) D(-\alpha)$ with the choice for the displacement $\alpha = 1.5$ for Hilbert-space dimension $N = 32$. The angles are chosen $\vec \theta = [\pi/2, \pi/2, -\pi/2, -\pi/2, \pi/2, \pi/2]$.


\subsection{Generating random processes}

The DV processes were generated by combining random unitaries $\{U^{\text{rand}}_i\}$ with random weights $w_i$ that add up to make $\{w_i U^{\text{rand}}_i\}$ a valid set of Kraus operators that sum up to $\mathbb I$. The unitaries are created by matrix exponentiation $U_i = e^{-iH_i}$ of random Hermitian matrices $H_i = 0.5(X_i + X_i^\dag)$. The matrices $X_i$ are random with elements generated by sampling uniformly from the range $[-1, 1]$.


\subsection{Neural-network quantum process tomography}
\label{app:nn-qpt}

In neural-network quantum process tomography (NN-QPT), we design a simple feed-forward neural network that takes as input batches of measurement data $d_{ij}$ and gives a set of $k$ Kraus operators as output. The structure of the network is described in \tabref{tab:nn}; it consists of a set of dense neurons with appropriate reshaping operations to generate the $k$ complex-valued Kraus operators. We use the hyperbolic tangent (tanh) as our activation function. Previous works have considered similar ideas for QST with a \texttt{DensityMatrix} layer shaping the output of a neural network to a valid density matrix~\cite{Ahmed2020a, Ahmed2020b}.

\begin{table}
\centering
\caption{The neural-network architecture used for QPT. An implementation of the model and its training procedure can be found in Ref.~\cite{gd-qpt2022}.
\label{tab:nn}}
\renewcommand{\arraystretch}{1.25}
\renewcommand{\tabcolsep}{0.15cm}
\begin{tabular}{|l|c|r|}
\hline
Layer & Output shape\\
\hline
Input (-) & \\
Dense & $128$ \\
Tanh & -  \\
Dense & $128 $\\
Tanh & - \\
Dense & $k \times N \times N \times 2$ \\
Reshape & $k \times N \times N$  \\
\hline
\end{tabular}
\end{table}

\begin{figure}
    \centering
    \includegraphics[width=\columnwidth]{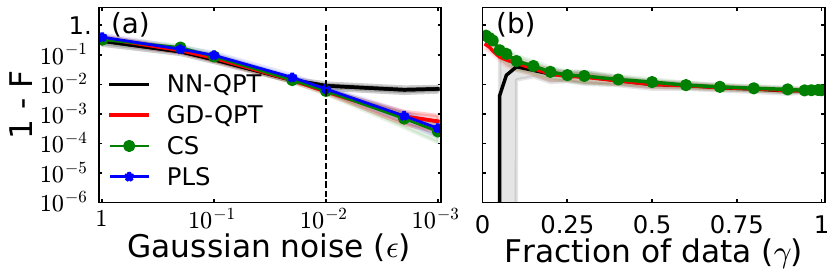}\\
    \includegraphics[width=\columnwidth]{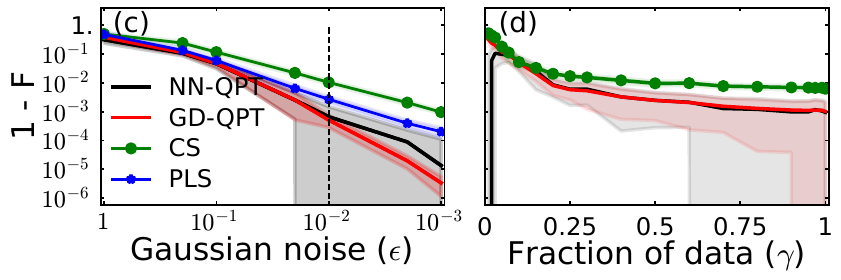}
    \caption{Neural-network QPT for random two-qubit ($n=2$) processes with different ranks using Pauli probes and measurements. (a, c) The mean infidelities as the Gaussian noise level $\epsilon$ is decreased for $30$ random processes with (a) $r=16$, $k=16$ and (c) $r=4$, $k=4$. (b, d) The mean infidelities for $30$ random process reconstructions using a fraction $\gamma$ of the total data used in (a, c). We randomly select a subset of the total $6^n \times 6^n$ Pauli probes and measurements, i.e., $\sqrt \gamma/6^n \times \sqrt \gamma/6^n$.
    \label{fig:nn-benchmark}}
\end{figure}

In \figref{fig:nn-benchmark}, we show benchmarks for NN-QPT against GD-QPT, PLS, and CS. The input to the model is batches of the data with batch size $16$ and the output is a set $\mathbb K^{\text{NN-QPT}}$ of Kraus operators for each batch. The measurements and probes for each batch along with $\mathbb K^{\text{NN-QPT}}$ gives data predictions $d_{ij}'$. The loss function is the least-squares error between $d_{ij}$ and $d_{ij}'$ with regularization and a penalty for violating the TP condition given as $0.001 \vert \vert \mathbb K - \mathbb I\vert \vert_1$. We use the Adam optimizer with a learning rate of $0.001$ during the training.

Once trained, the network generates a set of Kraus operators from a random batch of data as input. There is an ambiguity about which batch should be chosen to generate the final set of reconstructed Kraus operators. If all the data is used during training as well as prediction, this ambiguity is removed. Nevertheless, we find that all random sets of data as inputs gives similar output for the Kraus operators.

We also find that the reconstructed set of Kraus operators do not strictly follow the TP condition as the optimization only penalizes violations of the trace-preserving property instead of enforcing it. In future works, it would be interesting to combine the constrained-gradient approach with neural-network and other types of ansatzes such as tensor networks similar to recent attempts~\cite{Srinivasan2021} in order to ensure that the reconstructed process is TP.

In future works, the NN-QPT idea can be explored to determine how such neural-network approaches can achieve better performance than simple GD-QPT, or can be extended to larger system sizes without requiring a full Kraus representation. As in the case of GD-QPT, future work with NN-QPT would also require strict guarantees on the TP condition as well as uncertainity estimates to be useful in realistic experiments.

\end{document}